\documentclass[runningheads]{llncs}
\usepackage[T1]{fontenc}
\usepackage[unicode]{hyperref}
\usepackage[english]{babel}
\addto\extrasenglish{}
\addto\extrasenglish{}
\addto\extrasenglish{}
\addto\extrasenglish{}
\addto\extrasenglish{}
\addto\extrasenglish{}
\usepackage{acronym}
\usepackage{bm}
\usepackage{graphicx}
\usepackage{orcidlink}
\usepackage{algorithmic}
\usepackage{algorithm}
\usepackage{array}
\usepackage[caption=false,font=normalsize,labelfont=sf,textfont=sf]{subfig}
\usepackage{textcomp}
\usepackage{stfloats}
\usepackage{url}
\usepackage{verbatim}
\usepackage{graphicx}
\usepackage{cite}
 \usepackage{multirow}
 \usepackage{array, makecell} %
\usepackage{svg}
\usepackage{amssymb}

\usepackage{censor}
\acrodef{BLSTM}{bi-di\-rec\-tional long short-term memory}
\acrodef{CNN}{convolutional neural network}
\acrodef{GAN}  {generative adversarial network}
\acrodef{UDASE} {unsupervised domain adaptation speech enhancement}
\acrodef{CMGAN} {conformer-based metric \ac{GAN}}
\acrodef{SSSR}{self-supervised speech representation}
\acrodef{STOI}{Short-Time Objective Intelligibilty}
\acrodef{STFT}{short-time Fourier transform}
\acrodef{ISTFT}{inverse short-time Fourier transform}
\acrodef{PESQ}{Perceptual Evaluation of \ac{SQ}}
\acrodef{DNSMOS}{Deep Noise Suppression Mean Opinion Score}
\acrodef{MOS}{mean opinion score}
\acrodef{NISQA}{Non-Intrusive \ac{SQ} Assessment}
\acrodef{MUSHRA}{Multiple Stimuli with Hidden Reference and Anchor}
\acrodef{ASR}{automatic speech recognition}
\acrodef{RMSE}{Root Mean Squared Error}
\acrodef{MSE}{Mean Squared Error}
\acrodef{SE}{speech enhancement}
\acrodef{SI-SDR}{scale-invariant signal-distortion ratio}
\acrodef{SQ}{speech quality}
\acrodef{GELU}{Gaussian Error Linear Unit}

\begin{document}
\title{WhiSQA: Non-Intrusive Speech Quality Prediction Using Whisper Encoder Features}
%
%
\author{George Close$^{1}$,
    Kris Hong$^{1}$,
      Thomas Hain$^{2}$,
      and~Stefan~Goetze$^{2,3}$
      \thanks{This work was supported by the Centre for Doctoral Training in Speech and Language Technologies (SLT) and their Applications funded by UK Research and Innovation [grant number EP/S023062/1].}}
\institute{$^{1}$ConnexAI, Manchester, UK,\\
$^{2}$School of Computer Science, The University of Sheffield, Sheffield, UK,\\ 
$^{3}$South Westphalia University of Applied Sciences, Iserlohn, Germany 
\email{\{george.close,kris.hong\}@connex.ai, t.hain@sheffield.ac.uk, goetze.stefan@fh-swf.de    }
}
\authorrunning{G. Close et al.}
\titlerunning{WhiSQA}
%
%
\maketitle              
\begin{abstract}
There has been significant research effort developing neural-network-based predictors of \ac{SQ} in recent years. While a primary objective has been to develop non-intrusive, i.e.~reference-free, metrics to assess the performance of \ac{SE} systems, recent work has also investigated the direct inference of neural \ac{SQ} predictors within the loss function of downstream speech tasks. To aid in the training of \ac{SQ} predictors, several large datasets of audio with corresponding human labels of quality have been created. Recent work in this area has shown that speech representations derived from large unsupervised or semi-supervised foundational speech models are useful input feature representations for neural \ac{SQ} prediction. In this work, a novel and robust \ac{SQ} predictor is proposed based on feature representations extracted from an \ac{ASR} model, found to be a powerful input feature for the \ac{SQ} prediction task. The proposed system achieves higher correlation with human \ac{MOS} ratings than recent approaches on all NISQA test sets and shows significantly better domain adaption compared to the commonly used DNSMOS metric.
\end{abstract}

\section{Introduction}
To assess the performance of \acf{SE} methods, there is a continuing interest in the development of metrics to assess the \acf{SQ} of given input audio~\cite{moller2011speechQualityTrends,RGHKK08,loizou2013speech,GARHK13,NISQA,kumar2023torchaudiosquim}. Such metrics allow for the automatic assessment and comparison of \ac{SE} systems without the need for expensive and time-consuming human listening tests~\cite{MOS-ITU,mushra,GoetzeIWAENC14sq-si,Avila2016QualityComparison}. Many still commonly used metrics, such as the \ac{PESQ}~\cite{pesq} or \ac{STOI}~\cite{stoi} are signal-processing-based \emph{intrusive} metrics, i.e.~are designed to operate over an input of clean reference audio and a \emph{(typically artificially) corrupted} or \emph{enhanced} version of that same audio, the latter being the signal under assessment. From a neural network perspective, intrusive metrics based on traditional signal processing have two major drawbacks. Firstly, many traditional metrics have stages to their computation which cannot be easily formulated in a differentiable way, which renders them difficult to optimise towards within a loss function for neural-network-based \ac{SE} systems\cite{diff_pesq}. This limitation can partially be overcome by frameworks like MetricGAN~\cite{fu21_interspeech,close2022,cmgan,close2024multicmgan,mgk2025}, where an \ac{SE} network and a neural \emph{metric predictor} network are adversarially trained in a \ac{GAN} setting, but such networks might be prone to artifacts not properly assessed by the metric prediction~\cite{deoliveira24_interspeech,Close_hallucinations_EUSIPCO24}. The second major drawback of most traditional metrics is their intrusive nature; the reliance on the existence of the reference signal usually requires that test data be \textit{simulated} (i.e.~as artificially corrupted versions of the reference audio) rather than \emph{real} (i.e.~gathered in the 'real world' from the target domain of the system under test).\\
To overcome these drawbacks, several datasets and network structures~\cite{CSS+19,NISQA,reddy2022dnsmos,yi2022conferencingspeech,dong2020pyramid} for the task of neural non-intrusive \ac{SQ} prediction have been proposed. Datasets for the \ac{SQ} prediction task typically consist of noisy audio with associated human \ac{MOS}~\cite{MOS-ITU} quality labels that have been collected in listening tests conducted by human listeners. Neural networks can be trained with the noisy audio as input to predict the associated \ac{MOS} label.

In parallel with the \ac{SQ} prediction task is the related task of non-intrusive \textit{intelligibility} prediction~\cite{stoi,WKJ+14,ASR-based_SI_prediction2022,barker24_icassp}. As the datasets for this task are significantly smaller, much of the focus in this topic has been on finding powerful input feature representations rather than on designing large complex network structures. In particular, features derived from large, pre-trained \emph{foundation models} have shown to be particularly useful for the intelligibility prediction task~\cite{close2023non,winnerCPC2paper,mogridge2024nonintrusive}.

In this work, feature representations generated by a foundational model are analysed as input to a neural network for the \ac{SQ} prediction task. Such features, which have primarily been developed as backbone models for \ac{ASR} have proved to be useful feature representations for a number of speech related tasks~\cite{PasadLivescu23_SSSRAnalysis,close2023perceive}.  Experiments investigating different combinations of training data corpora with different score distributions are carried out, and the effects on test time performance are analysed. Although non-intrusive \ac{SQ} prediction is the main aim of this work, the identified best-performing models are analysed as intrusive and multi-headed (i.e.~predicting multiple labels at once) variants. State-of-the-art performance is achieved on common testsets using the proposed model. The implementation of the best performing model as a \ac{SQ} metric is provided online\footnote{available at \url{https://github.com/leto19/WhiSQA}}. 

The remainder of this work is structured as follows: \autoref{sec:Foundation-Model-Features} introduces the foundation model from which input feature representations for the model structure are extracted.  \autoref{Speech-Quality-(SQ)-Prediction-Models} formally introduces the \ac{SQ} prediction task and the proposed model structure. \autoref{Datasets-for-Speech-Quality-(SQ)-Prediction}  describes and analyses the \ac{SQ} datasets which are used to train, validate and test the proposed model. \autoref{sect:exp2}  details experiments in which the optimal training data setup and task variants are investigated before \autoref{sec:Conclusion} concludes the paper.

\section{Whisper  Features}\label{sec:Foundation-Model-Features}
Whisper is a weakly supervised Transformer-based \ac{ASR} system. It has shown state-of-the-art performance on a number of monolingual \ac{ASR} benchmark \linebreak datasets, as well as multilingual transcription and translation tasks~\cite{whisper}.

It consists of several sequential Transformer-based encoder blocks $\mathcal{A}_\mathrm{E}(\cdot)$ followed by the same number of sequential Transformer-based decoder blocks $\mathcal{A}_\mathrm{D}(\cdot)$. The input to the encoder $\mathcal{A}_\mathrm{E}(\cdot)$ is a log-Mel spectrogram representation $\mathbf{X}_\mathrm{MEL}$ of the input audio $x[n]$ (padded to $30$ seconds in length), which is processed by a $1$-dimensional \ac{CNN} layer and a~\ac{GELU} activation function, followed by a sinusoidal positional encoding before being processed by the first encoder Transformer block. The output of each encoder layer $\ell$ is denoted as $\mathbf{X}_\mathrm{E}^{(\ell)}$, a two-dimensional representation of dimension $768$ by $1500$~\cite{whisper}. 
The Whisper decoder $\mathcal{A}_\mathrm{D}(\cdot)$ takes the form of a language model; the first Transformer block of the decoder takes as input a sequence of tokens which encode the language, task, timestamp in seconds, and the previously transcribed words of the utterance. Each Transformer block in the decoder has access to the output of the encoder via a cross-attention mechanism. The final output of the decoder (not used in this work) is a prediction of the next token (i.e.~the next word) in the input sequence. The $T$ dimension of the output of each Whisper decoder layer is significantly smaller than any other feature used in this work. 

In this work, the \texttt{whisper-small}\footnote{\url{https://huggingface.co/openai/whisper-small}} model, trained on $680$k hours of labelled speech data is used. Recent work has found that features extracted from both the encoder~\cite{winnerCPC2paper} and decoder~\cite{mogridge2024nonintrusive} layers of Whisper are useful for capturing \emph{intelligibility}-related information. Hence, this work analyses their capability for \emph{quality} prediction.  
The encoder $\mathcal{A}_\mathrm{E}(\cdot)$ and decoder $\mathcal{A}_\mathrm{D}(\cdot)$ of this model each have $12$ transformer blocks; the set of outputs of each of the constituent transformer blocks are thus denoted as $\{\mathbf{X}_\mathrm{E}^{(0)}, ..., \mathbf{X}_\mathrm{E}^{(12)}\}$  and $\{\mathbf{X}_\mathrm{D}^{(0)}, ..., \mathbf{X}_\mathrm{D}^{(11)}\}$, respectively.
The weighted sum of  $\{\mathbf{X}_\mathrm{E}^{(0)}..\mathbf{X}_\mathrm{E}^{(12)}\}$ is defined as
\begin{equation}
    \mathbf{\bar{X}}_\mathrm{E} = \sum_{\ell=0}^{12} \alpha_\mathrm{E}^{(\ell)} \cdot \mathbf{X}_\mathrm{E}^{(\ell)},
    \label{eq:featureAveraging}
\end{equation}
where $\{\alpha_\mathrm{E}^{(0)},..,\alpha_\mathrm{E}^{(12)}\}$ are parameter weights for each layer which are learned during prediction model training. 
\section{Speech Quality (SQ) Prediction Models}
\label{Speech-Quality-(SQ)-Prediction-Models}
\label{sec:Speech-Quality-Prediction-Task}
\begin{figure}[!ht]
 \centering
 \resizebox{.85\columnwidth}{!}{%
 \graphicspath{{figs}} 
 \includesvg{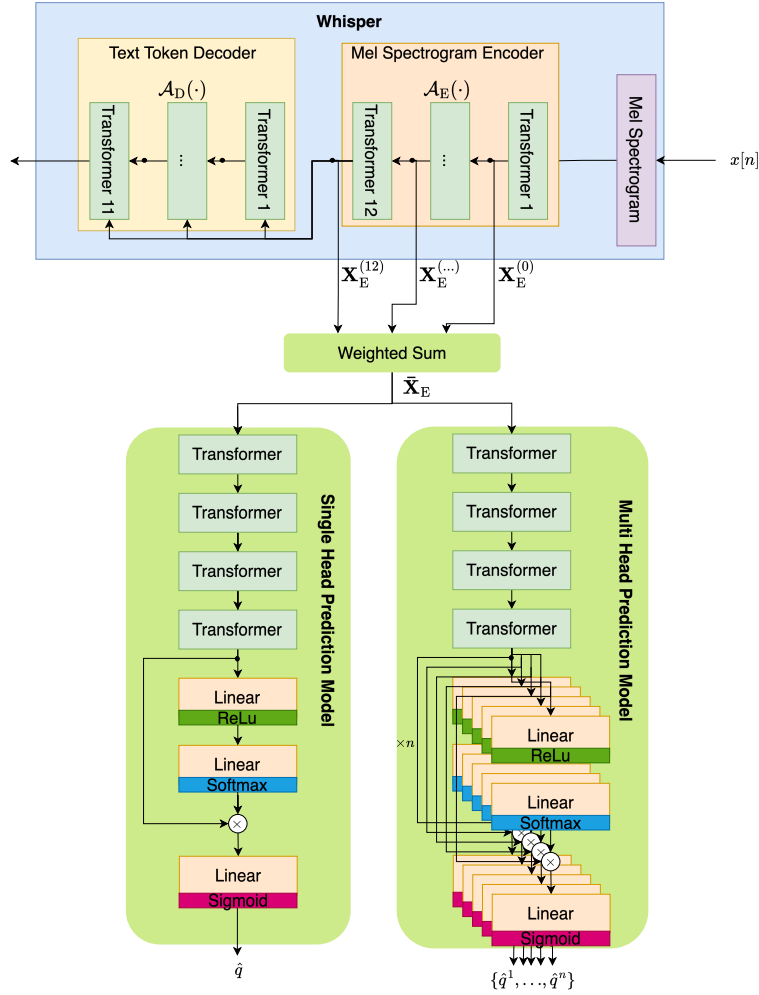} 
 }
\caption{Network structure of the proposed WhiSQA \ac{SQ} predictor with Whisper Encoder feature extraction. Note that the 'Weighted Sum' block contains model parameters, i.e.~layer weights $\{\alpha^{(0)},..,\alpha^{(12)}\}$ from (\ref{eq:featureAveraging}) which are updated during prediction model training.}
\label{fig:model_struct_new}
\end{figure}

For non-intrusive speech quality prediction, the neural network $\mathcal{D}(\cdot)$ takes as input a feature representation 
\begin{equation}
    \mathbf{X}_{\mathrm{F}}=\mathcal{F}(x[n])
    \label{eq:feature_extract}
\end{equation}
\noindent of the speech or audio signal under test $x[n]$ and returns a predicted quality label $\hat{q}$. The operator $\mathcal{F}(\cdot)$ denotes the feature extraction process; for this work $\bar{\mathbf{X}}_\mathrm{E}$ is taken as input features.
Typically, $\mathcal{D}(\cdot)$ is trained on data consisting of tuples $(x[n],q)$ where $q$ is the \emph{true} \ac{MOS} quality label of the audio $x[n]$ obtained from signal assessment by human listeners. The loss function used to train $\mathcal{D}(\cdot)$ is often a simple \ac{MSE} between the model output i.e the \emph{predicted} score $\hat{q}=\mathcal{D}(\mathbf{X}_{\mathrm{F}})$ and the \emph{true} quality label $q$:
\begin{equation}
    L_\mathcal{D} = (\mathcal{D}(\mathbf{X}_{\mathrm{F}}) - q)^2.
\end{equation}
Note that while \ac{MOS} labels are typically expressed in the range $1$ to $5$, higher being better, for the ease of training of neural \ac{SQ} predictors, $q$ is typically normalised to a range between $0.2$ and $1$, which enables a sigmoid activation function on the final neural network layer to project to this label range~\cite{fu2021metricganu}. \ac{SQ} prediction models can be broadly classified into two types; \emph{single-headed} models which predict only the \ac{MOS} label and \emph{multi-headed} models which predict \ac{MOS} alongside some other label(s) of the input audio, e.g.~Noisiness, Coloration, Discontinuity, etc. 
\\
The structure of the proposed \ac{SQ} prediction models $\mathcal{D}(\cdot)$ is based on \cite{bas-xlsr}, and is shown in \autoref{fig:model_struct_new}.
The  model $\mathcal{D}_1(\cdot)$ (denoted as `Single Head Prediction Model' in \autoref{fig:model_struct_new}) consists of $4$ transformer layers, followed by an attention pooling mechanism with a sigmoid activation function, which returns the predicted \ac{MOS} score $\hat{q}$ normalised between $0.2$ and $1$. The input dimension (and thus the parameter count) of the transformer stage depends on the feature dimension $F$ of the input feature, while the output dimension is fixed at $256$. The attention pooling  mechanism consists of two sequential linear layers, with a softmax function applied at the output and is multiplied by the output of the Transformer block. The result of this multiplication is further fed into a final linear layer with a sigmoid activation to a single output neuron. This single output neuron represents the predicted \ac{MOS} label $\hat{q}$ of the input audio.
A variant of this base model (denoted as `Multi Head Prediction Model' in \autoref{fig:model_struct_new}) which incorporates multiple prediction 'heads' i.e.~the three Linear layer structure is also proposed for multi-dimension speech quality prediction. 

\section{Datasets for Speech Quality Prediction}
\label{Datasets-for-Speech-Quality-(SQ)-Prediction}

Datasets containing \acf{MOS} scores $q$ obtained from  listening test with humans for signals under test $x[n]$ have only been created during the last few years in quantities which allow training recent data-driven methods. Several \ac{SQ} datasets are now available and briefly analysed in the following. It is important to consider several  datasets to ensure that the \ac{SQ} predictor has been exposed to a large variety of audio conditions during its training.
For some datasets and subsets within datasets, further information is available such as a clean reference signal $s[n]$, the standard deviation of the \ac{MOS} score, the raw scores assigned by each human evaluator or the number of human assessors. 

\subsection{NISQA Dataset}
The \ac{NISQA}~\cite{NISQA} dataset is an  \ac{SQ} assessment dataset, comprising of pre-defined train, validation and test sets. Each of these are further divided into subsets, characterised by if the nature of the distortion in the speech signal is artificially simulated or occurring 'in the wild' as a real distortion. In addition to \ac{MOS} scores of overall audio quality, the NISQA dataset also provides labels for other speech `dimensions'\cite{Wltermann2013DimensionbasedQM} namely Noisiness, Coloration, Discontinuity and Loudness. It has three defined testsets, denoted as FOR, LIVETALK and P501. With the exception of the LIVETALK testset, clean reference signals $x[n]$ are available.  
The baseline \ac{NISQA} model has single and multi-headed variants.

\subsection{Tencent Dataset}
The Tencent audio \ac{SQ} dataset was released as part of the ConferencingSpeech 2022 challenge~\cite{yi2022conferencingspeech}. It consists of two artificially simulated training subsets, one with artificial reverberation added and one without. 

\subsection{Indiana University Bloomington (IUB) Dataset}
The Indiana University Bloomington (IUB)~\cite{dong2020pyramid} \ac{SQ} dataset consists of two subsets. The first uses distorted audio sourced from the COnversational Speech In Noisy Environments (COSINE)~\cite{acoustics3010014} dataset, real multi party conversations captured using multi-channel wearable microphones recorded in noisy everyday environments. The second subset uses audio from the Voices Obscured in Complex Environmental Settings (VOiCES)\cite{VOICES} corpus where speech and noise were played aloud and recorded in two rooms of different sizes.\\ Unlike the other datasets used in this work, the \ac{MOS} scores for this dataset were gathered using a \ac{MUSHRA}\cite{mushra} protocol, which is then transformed to a \ac{MOS} scale between $0$ and $10$, rather than the $1$ to $5$ scale commonly used. The $1$ - $5$ MOS label is obtained via a fitting operation over the gathered MUSHRA ratings. 

\subsection{Public Switched Telephone Network (PSTN) Dataset}
The Public Switched Telephone Network (PSTN) \ac{SQ} dataset~\cite{mittag20b_interspeech} consists of simulated 'real' phone calls, some with simulated background noise added to the transmitted signal. It follows a similar design to that of NISQA, but is significantly larger. 

\subsection{Overall MOS Distribution}
\begin{figure}[!h]
    \centering
    \resizebox{0.8\columnwidth}{!}{
    \includesvg{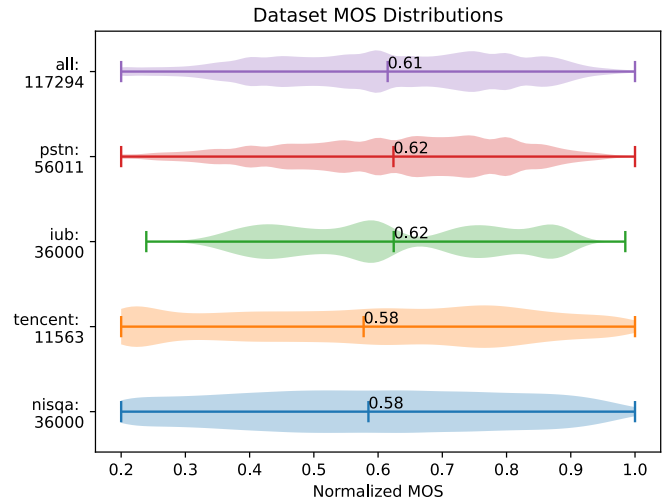}
    }
    \caption{Normalised MOS score distribution across \ac{SQ} datasets with lines indicating minimum, mean and maximum \ac{MOS} in each dataset.Numbers on y axis denote number of data points in each set.}
    \label{fig:mos-violin}
\end{figure}

To compare the available datasets and analyse prediction results later in this paper, the distributions of \ac{MOS} scores in the training and validation subsets of the datasets (normalised between $0.2$ and $1$) are shown in \autoref{fig:mos-violin}. The mean \ac{MOS} value across the datasets is similar, at approximately $0.65$. However, the datasets differ significantly in the shape of their distributions. Both \ac{NISQA} and Tencent show a roughly uniform distribution of scores from $0.2$ to $1$, with the `tail' at the lower end of the Tencent distribution showing that that dataset contains a larger numbers of low scores. Conversely, the tapering in at the highest end in both NISQA and Tencent indicate that these datasets contain relatively few instances of very highly rated audio. 

In contrast, the distribution of the PSTN dataset scores is generally normal, tailing off at the low and high end. Slightly more scores are above $0.5$ than below, indicating that the audio in this dataset is generally of high quality. 

The distribution of the \ac{MOS} score in the IUB dataset is most different from the others, with very few points falling a the highest and lowest values. Further, it is significantly more erratic than the other datasets, with an extreme dearth in scores valued around $0.65$. This can possibly be explained by the non-standard method that the MOS scores were gathered, as well as the differing range of the scores before normalisation.

The combined distribution across all the datasets is shown in purple at the top of \autoref{fig:mos-violin}. It displays a similar normal-like distribution to that of the PSTN dataset, likely due to that dataset contributing roughly half of all samples. There are slightly more samples of low quality compared to high quality.
\section{Experiments}
\label{sect:exp2}
This experiment aims to find which training datasets have the greatest effect on test performance of the proposed \ac{SQ} prediction networks, as well as enabling a fair comparison with other recently proposed \ac{SQ} prediction systems. 
\subsection{Experiment Setup}
All models are  tested on each of the three NISQA test sets, i.e.~FOR, LIVETALK and P501. Following \cite{NISQA}, a training strategy where training stops only if the validation performance does not improve after $20$ epochs is employed. The bias-aware loss function, scaling the contribution of the training samples in the loss computation based on the relative size of the training set/subset, as proposed in \cite{NISQA} is also used here. The Adam~\cite{kingma2017adam} optimiser is used with an initial learning rate of $0.00001$, which is reduced by a factor of $0.1$ if the validation loss does not improve after $15$ epochs. All models are at first trained over a warmup epoch, where the learning rate increases up to the initial learning rate after each model update. A batch size $B$ of $128$ is used. The best-performing epoch on the validation set in terms of validation loss is loaded at test time. Datasets other than NISQA 
do not have defined validation sets; for these, $10\%$ of the training sets are partitioned for validation, following \cite{msqat}. All possible permutations of the evaluated datasets are used. The proposed Multi Head model (right in \autoref{fig:model_struct_new}) is trained on the NISQA testset to predict the \ac{MOS} as well as the Noisiness, Coloration, Discontinuity and Loudness
labels.

Models are evaluated using Spearman correlation $r$ and \ac{MSE} $e$, computed versus the true \ac{MOS} value for each testset element. 

\subsection{Results}
\begin{table}[!ht]
\centering
\caption{Training Data Ablation Study for best performing proposed single-head model. \textbf{Best} and \underline{second best} shown in \textbf{bold} and \underline{underlined}, respectively.}
\label{tab:dataset-ablation}
\resizebox{\textwidth}{!}{%
\begin{tabular}{ccccc|cc|cc|cc|cc}
\multicolumn{4}{l}{\textbf{Training Data}}          & \multicolumn{1}{r|}{} & \multicolumn{2}{c|}{\textbf{FOR}}                 & \multicolumn{2}{c|}{\textbf{LIVETALK}}            & \multicolumn{2}{c|}{\textbf{P501}}                & \multicolumn{2}{c}{\textbf{AVERAGE}}             \\
NISQA& Tencent  & IUB & PSTN                        & \textbf{Train Points}              & {r $\uparrow$} & \textbf{$e$}$\downarrow$ & {r $\uparrow$} & \textbf{$e$} $\downarrow$ & {r $\uparrow$} & \textbf{$e$} $\downarrow$ & {r $\uparrow$} & \textbf{$e$} $\downarrow$ \\ \hline
        &    \checkmark   &     & 
                     & \phantom{0}9250                                  & 0.82                  & 0.50                      & 0.83                  & 0.56                      & 0.83                  & 0.56                      & 0.83                  & 0.54                      \\
     \checkmark   &       &     &                        & 11020                                 & 0.92                  & 0.35                      & 0.82                  & 0.54                      & 0.93                  & 0.37                      & 0.89                  & 0.44                      \\
     \checkmark   &\checkmark       &     &              & 20270                                 & 0.93                  & 0.32                      & 0.87            & 0.46                      & \textbf{0.93}         & \underline{ 0.37}                & \underline{ 0.91}            & \underline{ 0.38}                \\
        &       &   \checkmark  &                         & 28800                                 & 0.27                  & 0.84                      & 0.42                  & 0.85                      & 0.41                  & 0.92                      & 0.37                  & 0.87                      \\
        &     \checkmark  &  \checkmark   &                & 38050                                 & 0.85                  & 0.46                      & 0.76                  & 0.62                      & 0.79                  & 0.62                      & 0.80                  & 0.57                      \\
     \checkmark   &       &  \checkmark   &                  & 39820                                 & 0.93                  & 0.32                      & 0.83                  & 0.52                      & 0.92                  & 0.40                      & 0.89                  & 0.41                      \\
        &       &    \checkmark &                         & 44809                                 & 0.92                  & 0.34                      & 0.77                  & 0.60                      & 0.88                  & 0.48                      & 0.86                  & 0.47                      \\
     \checkmark   &  \checkmark     &  \checkmark   &        & 49070                                 & 0.93                  & 0.32                      & 0.86                  & 0.48                      & 0.91                  & 0.42                      & 0.90                  & 0.41                      \\
        &    \checkmark   &     & \checkmark              & 54059                                 & 0.91                  & 0.36                      & 0.85                  & \textbf{0.39}             & 0.90                  & 0.45                      & 0.89                  & 0.40                      \\
    \checkmark    &       &     & \checkmark                & 55829                                 & \textbf{0.94}         & \textbf{0.29}             & 0.83                  & 0.51                      & \textbf{0.94}         & \textbf{0.35}             & 0.90                  & 0.38                      \\
    \checkmark    &  \checkmark     &     & \checkmark       & 65079                                 & \underline{ 0.94}            & \underline{ 0.30}                & \textbf{0.88}         & 0.45                & 0.93                  & 0.38                      & \textbf{0.92}         & \textbf{0.38}             \\
        &       &  \checkmark   & \checkmark               & 73609                                 & 0.89                  & 0.40                      & 0.72                  & 0.65                      & 0.76                  & 0.39                      & 0.79                  & 0.48                      \\
        &    \checkmark   &  \checkmark   &\checkmark        & 82859                                 & 0.92                  & 0.34                      & 0.81                  & 0.55                      & 0.83                  & 0.56                      & 0.85                  & 0.48                      \\
    \checkmark    &       &\checkmark     & \checkmark          & 84629                                 & 0.94                  & 0.30                      & 0.87                  & 0.46                      & 0.93                  & 0.39                      & 0.91                  & 0.38                      \\
       \checkmark &    \checkmark   & \checkmark    &\checkmark & 93879                                 & 0.93                  & 0.31                      & \underline{0.88}                  & \underline{0.45}                      & 0.91                  & 0.42                      & 0.91                  & 0.39                     
\end{tabular}
}
\end{table}
\autoref{tab:dataset-ablation} shows the results for the training data ablation experiment for the three NISQA test sets. The overall (on average) best-performing combination of training datasets is "NISQA, Tencent and PSTN".  
By far the lowest-performing model is that trained solely on IUB; further, also any given combination of training datasets including IUB performs worse on average than that combination without IUB. 
As noted earlier in \autoref{Datasets-for-Speech-Quality-(SQ)-Prediction}, this is likely due to the significantly different distribution of the MOS labels in this dataset relative to the others. 
The overall size of the training set has a smaller effect on performance - the inclusion of data more similar to the test sets (i.e.~the NISQA training data) results in better performance. This can perhaps be attributed to the bias-aware loss function used, which attempts to control for the imbalance in size between the component datasets.  
It can be noted, that including the Chinese-language Tencent dataset in training generally improves performance on the German-language LIVETALK testset; this can be attributed to these models being better able to generalise to languages other than English.

\autoref{tab:sota-comp} shows a comparison the proposed system with three state-of-the-art neural \ac{SQ} predictor systems~\cite{NISQA,bas-xlsr,msqat}. Results for the proposed system trained on the same combination of data are shown for a fair comparison. For all training data combinations, the proposed WhiSQA system outperforms the SOTA system.

\begin{table}[!htb]
\centering
\caption{Comparison of WhisSQA with SOTA systems. \textbf{Best} and \underline{second best} shown in \textbf{bold} and \underline{underlined}, respectively.}\label{tab:sota-comp}
\resizebox{\textwidth}{!}{%
\begin{tabular}{l|l|cc|cc|cc|cc}
                  & & \multicolumn{2}{c|}{\textbf{FOR}} & \multicolumn{2}{c|}{\textbf{LIVETALK}} & \multicolumn{2}{c|}{\textbf{P501}} & \multicolumn{2}{c}{\textbf{AVERAGE}} \\ \hline
    \textbf{Model }& \textbf{Training Data }& {r $\uparrow$} & \textbf{$e$} $\downarrow$ & {r $\uparrow$} & \textbf{$e$} $\downarrow$ & {r $\uparrow$} & \textbf{$e$} $\downarrow$ & {r $\uparrow$} & \textbf{$e$} $\downarrow$ \\
\textit{NISQA Single Head}~\cite{NISQA}            & \textit{NISQA}            & \textit{0.88} & \textit{0.40}  & \textit{0.70} & \textit{0.67}     & \textit{0.89} & \textit{0.46}                     & \textit{0.82} & \textit{0.51}            \\
Proposed WhiSQA & NISQA & \underline{0.92}                                  & \underline{0.35}             & 0.82                                  & 0.54             & \underline{0.93}           & \underline{0.37}              & 0.89             & 0.44 \\
MSQAT~\cite{msqat} & NISQA + Tencent + PSTN &  0.90 & 0.39 & 0.85 & 0.51  & 0.92 & 0.42 & 0.89 & 0.44\\
Proposed WhiSQA & NISQA+ Tencent + PSTN      & \textbf{0.94                                  }& \textbf{0.30}          & \textbf{0.88}                        & 0.45    & \textbf{0.93}  & \textbf{0.38}    & \textbf{0.92}    & \textbf{0.38}      \\
XLS-R SQA~\cite{bas-xlsr} &  Tencent + PSTN & 0.90 &0.38 & 0.83 & 0.52 & 0.89 & 0.46 & 0.82 & 0.51 \\
Proposed WhiSQA & Tencent + PSTN &    0.91&	0.36&	\underline{0.85}	&\textbf{0.39}&	0.90&	0.45&	\underline{0.89}	&\underline{0.40}                    \\
\end{tabular}
}
\end{table}
\autoref{tab:multi-head} compares the performance of the baseline NISQA model and the proposed model for multi-head / multi-label prediction. In both cases, the proposed system outperforms the NISQA baselines. For both systems, tasking the model with additionally predicting the other speech dimensions from the input audio slightly degrades the performance of the main task, i.e.~quality \ac{MOS} prediction. 
\begin{table}[!ht]
\centering
\caption{ \ac{MOS} prediction results for Multi Headed (MH) $\mathcal{D}_1$ Models versus Single Head (SH) Prediction. \textbf{Best} shown in \textbf{Bold}.}
\label{tab:multi-head}
\begin{tabular}{l|cc|cc|cc|cc}
 & \multicolumn{2}{c|}{\textbf{FOR}} & \multicolumn{2}{c|}{\textbf{LIVETALK}} & \multicolumn{2}{c|}{\textbf{P501}} & \multicolumn{2}{c}{\textbf{AVERAGE}} \\ \hline
\textbf{Model} & {r $\uparrow$} & \textbf{$e$} $\downarrow$ & {r $\uparrow$} & \textbf{$e$} $\downarrow$ & \textbf{r} $\uparrow $  & \textbf{$e$} $\downarrow$ & \textbf{r} $\uparrow $ & \textbf{$e$} $\downarrow$ \\
\textit{NISQA SH}  & \textit{0.88} & \textit{0.40} & \textit{0.70} & \textit{0.67} & \textit{0.89} & \textit{0.46} & \textit{0.82} & \textit{0.51}  \\
\textit{NISQA MH} & \textit{0.87} & \textit{0.43} & \textit{0.65} & \textit{0.72} & \textit{0.89} & \textit{0.46} & \textit{0.80} & \textit{0.54} \\
 WhiSQA SH & \textbf{0.92} & \textbf{0.35} & \textbf{0.82} & \textbf{0.54} & \textbf{0.93} & \textbf{0.37} & \textbf{0.89} & \textbf{0.42} \\
 WhiSQA MH & 0.91 & 0.36 & 0.69 & 0.58 & 0.92 & 0.41 & 0.84 & 0.45
\end{tabular}
\end{table}

\begin{figure}[!htb]
    \centering
    \resizebox{0.82\columnwidth}{!}{
    \includesvg{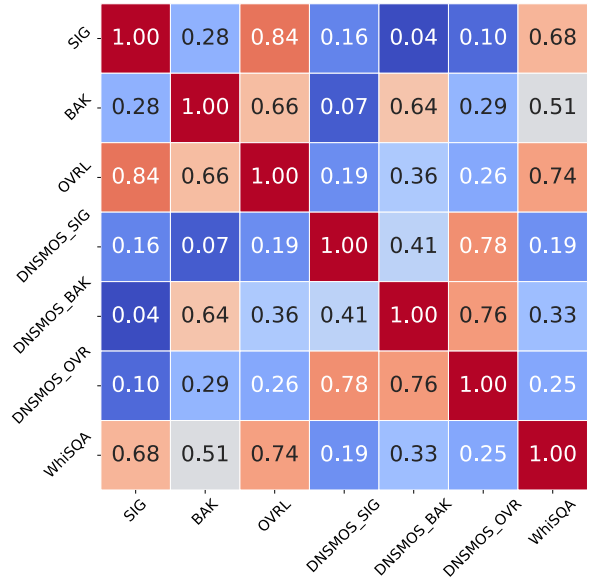}
    }
    \caption{Spearman Correlation Matrix for CHiME7-UDASE listening test data for DNSMOS and WhiSQA.}
    \label{fig:corr_mat}
\end{figure}
\autoref{fig:corr_mat} shows a Spearman correlation matrix for the CHiME7-\ac{UDASE} listening test~\cite{leglaive2023chime7}. This listening test was designed to assess the enhancement performance of the entries to the \ac{UDASE} challenge. \autoref{fig:corr_mat} compares human \ac{MOS} (\texttt{SIG}, \texttt{BAK} and \texttt{OVRL}) with those predicted by the DNSMOS~\cite{reddy2022dnsmos} metric (\texttt{DNSMOS\_SIG}, \texttt{DNSMOS\_BAK}, \texttt{DNSMOS\_OVRL}) and by the proposed single head WhiSQA model. \emph{The WhiSQA score correlates significantly more strongly} with the true \texttt{SIG} and \texttt{OVRL} scores compared to the corresponding DNSMOS metric value, while showing similar correlation to the true \texttt{BAK} score that the \texttt{DNSMOS\_BAK} metric does. 

\section{Conclusion and Future Work}
\label{sec:Conclusion}
This work introduces WhiSQA, a new SOTA system for speech quality prediction, as single- and multi-headed variants. Alayses for different datasets show improved performance over several baselines. Future work will explore further refinement of the system in the form of adaption to online `in the wild' data  as well as the applications of the Whisper encoder feature to other audio classification and evaluation tasks. 
%
%
%
%
\bibliographystyle{IEEEtran}
\bibliography{refs}

\begin{thebibliography}{10}
\providecommand{\url}[1]{#1}
\csname url@samestyle\endcsname
\providecommand{\newblock}{\relax}
\providecommand{\bibinfo}[2]{#2}
\providecommand{\BIBentrySTDinterwordspacing}{\spaceskip=0pt\relax}
\providecommand{\BIBentryALTinterwordstretchfactor}{4}
\providecommand{\BIBentryALTinterwordspacing}{\spaceskip=\fontdimen2\font plus
\BIBentryALTinterwordstretchfactor\fontdimen3\font minus \fontdimen4\font\relax}
\providecommand{\BIBforeignlanguage}[2]{{%
\expandafter\ifx\csname l@#1\endcsname\relax
\typeout{** WARNING: IEEEtran.bst: No hyphenation pattern has been}%
\typeout{** loaded for the language `#1'. Using the pattern for}%
\typeout{** the default language instead.}%
\else
\language=\csname l@#1\endcsname
\fi
#2}}
\providecommand{\BIBdecl}{\relax}
\BIBdecl

\bibitem{moller2011speechQualityTrends}
S.~M{\"o}ller, W.-Y. Chan, N.~C{\^o}t{\'e}, T.~H. Falk, A.~Raake, and M.~W{\"a}ltermann, ``Speech quality estimation: Models and trends,'' \emph{IEEE Signal Processing Magazine}, vol.~28, no.~6, pp. 18--28, 2011.

\bibitem{RGHKK08}
T.~Rohdenburg, S.~Goetze, V.~Hohmann, K.-D. Kammeyer, and B.~Kollmeier, ``{O}bjective {P}erceptual {Q}uality {A}ssessment for {S}elf-{S}teering {B}inaural {H}earing {A}id {M}icrophone {A}rrays,'' in \emph{Proc.\ IEEE Int.~Conf.~on Acoustics, Speech and Signal Processing (ICASSP)}, 2008.

\bibitem{loizou2013speech}
P.~Loizou, \emph{Speech Enhancement: Theory and Practice, Second Edition}.\hskip 1em plus 0.5em minus 0.4em\relax CRC Press, 2013.

\bibitem{GARHK13}
S.~Goetze, E.~Albertin, J.~Rennies, E.~Habets, and K.-D. Kammeyer, ``{S}peech {Q}uality {A}ssessment for {L}istening-{R}oom {C}ompensation,'' \emph{J. Audio Eng. Soc.}, vol.~62, no.~6, 2014.

\bibitem{NISQA}
G.~Mittag, B.~Naderi, A.~Chehadi, and S.~Möller, ``{NISQA}: A deep {CNN}-self-attention model for multidimensional speech quality prediction with crowdsourced datasets,'' in \emph{Interspeech 2021}, Aug. 2021.

\bibitem{kumar2023torchaudiosquim}
A.~Kumar, K.~Tan, Z.~Ni, P.~Manocha, X.~Zhang, E.~Henderson, and B.~Xu, ``Torchaudio-squim: Reference-less speech quality and intelligibility measures in torchaudio,'' 2023.

\bibitem{MOS-ITU}
{International Telecommunication Union}, ``{Recommendation {ITU-T P.800.2} {M}ean opinion score interpretation and reporting},'' {ITU}, ITU-T Recommendation, Jul. 2016.

\bibitem{mushra}
------, ``{Recommendation {ITU-R BS.1534-3} Method for the Subjective Assessment of Intermediate Quality Level of Audio Systems},'' {ITU}, ITU-R Recommendation, Oct. 2015.

\bibitem{GoetzeIWAENC14sq-si}
S.~Goetze, A.~Warzybok, I.~Kodrasi, J.~O. Jungmann, B.~Cauchi, J.~Rennies, E.~A.~P. Habets, A.~Mertins, T.~Gerkmann, S.~Doclo, and B.~Kollmeier, ``A study on speech quality and speech intelligibility measures for quality assessment of single-channel dereverberation algorithms,'' in \emph{Int.~Workshop on Acoustic Signal Enhancement (IWAENC)}, 2014.

\bibitem{Avila2016QualityComparison}
A.~Avila, B.~Cauchi, S.~Goetze, S.~Doclo, and T.~Falk, ``Performance comparison of intrusive and non-intrusive instrumental quality measures for enhanced speech,'' in \emph{Int.~Workshop on Acoustic Signal Enhancement (IWAENC)}, 2016.

\bibitem{pesq}
A.~Rix, J.~Beerends, M.~Hollier, and A.~Hekstra, ``Perceptual evaluation of speech quality (pesq)-a new method for speech quality assessment of telephone networks and codecs,'' in \emph{2001 IEEE ICASSP}, 2001.

\bibitem{stoi}
C.~H. Taal, R.~C. Hendriks, R.~Heusdens, and J.~Jensen, ``A short-time objective intelligibility measure for time-frequency weighted noisy speech,'' in \emph{ICASSP 2010}.

\bibitem{diff_pesq}
J.~Martín-Doñas, A.~Gomez, J.~Gonzalez~Lopez, and A.~Peinado, ``A deep learning loss function based on the perceptual evaluation of the speech quality,'' \emph{IEEE Signal Processing Letters}, vol.~PP, pp. 1--1, 09 2018.

\bibitem{fu21_interspeech}
S.-W. Fu, C.~Yu, T.-A. Hsieh, P.~Plantinga, M.~Ravanelli, X.~Lu, and Y.~Tsao, ``{MetricGAN+: An Improved Version of MetricGAN for Speech Enhancement},'' in \emph{Proc. Interspeech 2021}, 2021, pp. 201--205.

\bibitem{close2022}
G.~Close, T.~Hain, and S.~Goetze, ``{MetricGAN+/-: Increasing Robustness of Noise Reduction on Unseen Data},'' in \emph{EUSIPCO 2022}, Belgrade, Serbia, Aug. 2022.

\bibitem{cmgan}
R.~Cao, S.~Abdulatif, and B.~Yang, ``{CMGAN: Conformer-based Metric GAN for Speech Enhancement},'' in \emph{Proc. Interspeech 2022}, 2022, pp. 936--940.

\bibitem{close2024multicmgan}
G.~Close, W.~Ravenscroft, T.~Hain, and S.~Goetze, ``{Multi-CMGAN+/+: Leveraging Multi-Objective Speech Quality Metric Prediction for Speech Enhancement},'' in \emph{IEEE Int.~Conf.~on Acoustics, Speech and Signal Processing (ICASSP'24)}, 2024.

\bibitem{mgk2025}
Y.~Mai and S.~Goetze, ``{MetricGAN+KAN: Kolmogorov-Arnold Networks in Metric-Driven Speech Enhancement Systems},'' in \emph{Proc.\ Int.\ Conf.\ on Acoustics, Speech, and Signal Processing (ICASSP'25)}, 2025.

\bibitem{deoliveira24_interspeech}
D.~{de Oliveira}, S.~Welker, J.~Richter, and T.~Gerkmann, ``The pesqetarian: On the relevance of goodhart's law for speech enhancement,'' in \emph{Interspeech 2024}, 2024, pp. 3854--3858.

\bibitem{Close_hallucinations_EUSIPCO24}
G.~Close, T.~Hain, and S.~Goetze, ``Identifying hallucination in perceptually motivated speech enhancement networks,'' in \emph{32nd European Signal Processing Conference (EUSIPCO24)}, Lyon, France, Aug. 2024.

\bibitem{CSS+19}
B.~{Cauchi}, K.~{Siedenburg}, J.~F. {Santos}, T.~H. {Falk}, S.~{Doclo}, and S.~{Goetze}, ``{N}on-{I}ntrusive {S}peech {Q}uality {P}rediction {U}sing {M}odulation {E}nergies and {LSTM}-{N}etwork,'' \emph{IEEE/ACM Transactions on Audio, Speech, and Language Processing}, vol.~27, no.~7, Jul. 2019.

\bibitem{reddy2022dnsmos}
C.~K.~A. Reddy, V.~Gopal, and R.~Cutler, ``Dnsmos p.835: A non-intrusive perceptual objective speech quality metric to evaluate noise suppressors,'' 2022.

\bibitem{yi2022conferencingspeech}
G.~Yi, W.~Xiao, Y.~Xiao, B.~Naderi, S.~Möller, W.~Wardah, G.~Mittag, R.~Culter, Z.~Zhang, D.~S. Williamson, F.~Chen, F.~Yang, and S.~Shang, ``{ConferencingSpeech 2022 Challenge: Non-intrusive Objective Speech Quality Assessment (NISQA) Challenge for Online Conferencing Applications},'' in \emph{Proc. Interspeech 2022}, 2022, pp. 3308--3312.

\bibitem{dong2020pyramid}
X.~Dong and D.~S. Williamson, ``{A pyramid recurrent network for predicting crowdsourced speech-quality ratings of real-world signals},'' in \emph{Interspeech}, 2020, pp. 4631--4635.

\bibitem{WKJ+14}
A.~Warzybok, I.~Kodrasi, J.~Jungmann, E.~Habets, T.~Gerkmann, A.~Mertins, S.~Doclo, B.~Kollmeier, and S.~Goetze, ``Subjective speech quality and speech intelligibility evaluation of single-channel dereverberation algorithms,'' in \emph{Proc.~Int.~Workshop on Acoustic Signal Enhancement (IWAENC 2014)}, Sep. 2014.

\bibitem{ASR-based_SI_prediction2022}
M.~Karbasi and D.~Kolossa, ``Asr-based speech intelligibility prediction: A review,'' \emph{Hearing Research}, vol. 426, 2022.

\bibitem{barker24_icassp}
J.~Barker, M.~Akeroyd, W.~Bailey, T.~J. Cox, J.~F. Culling, J.~Firth, S.~Graetzer, and G.~Naylor, ``{The 2nd Clarity Prediction Challenge: A machine learning challenge for hearing aid intelligibility prediction},'' in \emph{ICASSP}, 2024.

\bibitem{close2023non}
G.~Close, T.~Hain, and S.~Goetze, ``Non intrusive intelligibility predictor for hearing impaired individuals using self supervised speech representations,'' in \emph{Proc.~Workshop on Speech Foundation Models and their Performance Benchmarks (SPARKS), ASRU sattelite workshop}, 2023.

\bibitem{winnerCPC2paper}
{Santiago Cuervo, Ricard Marxer}, ``{Temporal-hierarchical features from noise-robust speech foundation models for non-intrusive intelligibility prediction},'' in \emph{Clarity Workshop 2022}, 2022.

\bibitem{mogridge2024nonintrusive}
R.~Mogridge, G.~Close, R.~Sutherland, T.~Hain, J.~Barker, S.~Goetze, and A.~Ragni, ``Non-intrusive speech intelligibility prediction for hearing-impaired users using intermediate asr features and human memory models,'' in \emph{IEEE Int.~Conf.~on Acoustics, Speech and Signal Processing (ICASSP'24)}, 2024.

\bibitem{PasadLivescu23_SSSRAnalysis}
A.~Pasad, B.~Shi, and K.~Livescu, ``{Comparative Layer-Wise Analysis of Self-Supervised Speech Models},'' in \emph{IEEE Int.~Conf.~on Acoustics, Speech and Signal Processing (ICASSP)}, 2023.

\bibitem{close2023perceive}
G.~Close, W.~Ravenscroft, T.~Hain, and S.~Goetze, ``Perceive and predict: self-supervised speech representation based loss functions for speech enhancement,'' in \emph{Proc. ICASSP 2023}, 2023.

\bibitem{whisper}
A.~Radford, J.~W. Kim, T.~Xu, G.~Brockman, C.~McLeavey, and I.~Sutskever, ``{Robust Speech Recognition via Large-Scale Weak Supervision},'' 2022.

\bibitem{fu2021metricganu}
S.-W. Fu, C.~Yu, K.-H. Hung, M.~Ravanelli, and Y.~Tsao, ``Metricgan-u: Unsupervised speech enhancement/ dereverberation based only on noisy/ reverberated speech,'' 2021.

\bibitem{bas-xlsr}
B.~Tamm, R.~Vandenberghe, and H.~Van~hamme, ``Analysis of xls-r for speech quality assessment,'' in \emph{Proc. WASPAA 2023}, 10 2023, pp. 1--5.

\bibitem{Wltermann2013DimensionbasedQM}
\BIBentryALTinterwordspacing
M.~W{\"a}ltermann, ``Dimension-based quality modeling of transmitted speech,'' 2013. [Online]. Available: \url{https://api.semanticscholar.org/CorpusID:63687570}
\BIBentrySTDinterwordspacing

\bibitem{acoustics3010014}
\BIBentryALTinterwordspacing
A.~Hashmi, ``Perceptual evaluation of speech quality for inexpensive recording equipment,'' \emph{Acoustics}, vol.~3, no.~1, pp. 200--211, 2021. [Online]. Available: \url{https://www.mdpi.com/2624-599X/3/1/14}
\BIBentrySTDinterwordspacing

\bibitem{VOICES}
C.~Richey, M.~Barrios, Z.~Armstrong, C.~Bartels, H.~Franco, M.~Graciarena, A.~Lawson, M.~Nandwana, A.~Stauffer, J.~Hout, P.~Gamble, J.~Hetherly, C.~Stephenson, and K.~Ni, ``Voices obscured in complex environmental settings (voices) corpus,'' 04 2018.

\bibitem{mittag20b_interspeech}
G.~Mittag, R.~Cutler, Y.~Hosseinkashi, M.~Revow, S.~Srinivasan, N.~Chande, and R.~Aichner, ``{DNN No-Reference PSTN Speech Quality Prediction},'' in \emph{Proc. Interspeech 2020}, 2020.

\bibitem{kingma2017adam}
D.~P. Kingma and J.~Ba, ``Adam: A method for stochastic optimization,'' \emph{CoRR}, 2014.

\bibitem{msqat}
\BIBentryALTinterwordspacing
K.~Shen, D.~Yan, and L.~Dong, ``Msqat: A multi-dimension non-intrusive speech quality assessment transformer utilizing self-supervised representations,'' \emph{Applied Acoustics}, vol. 212, p. 109584, 2023. [Online]. Available: \url{https://www.sciencedirect.com/science/article/pii/S0003682X23003821}
\BIBentrySTDinterwordspacing

\bibitem{leglaive2023chime7}
S.~Leglaive, L.~Borne, E.~Tzinis, M.~Sadeghi, M.~Fraticelli, S.~Wisdom, M.~Pariente, D.~Pressnitzer, and J.~R. Hershey, ``{The CHiME-7 UDASE task: Unsupervised domain adaptation for conversational speech enhancement},'' 2023.

\end{thebibliography}
\end{document}